\documentclass[a4paper,11pt, nofootinbib]{article}
\pdfoutput=1 %
\usepackage{jcap} %
\usepackage{aas_macros}
\usepackage{newtxtext,newtxmath}

\usepackage[T1]{fontenc}
\usepackage{ae,aecompl}
\usepackage{graphicx}	%
\usepackage{amsmath}	%
\usepackage{amssymb}	%
\usepackage{multicol}	%
\usepackage{bm}		%
\usepackage{pdflscape}	%
\usepackage{hyperref}
\usepackage{ulem} %
\usepackage{lineno} %

\usepackage{graphicx}	%
\usepackage{amsmath}	%
\usepackage{amssymb}	%

\newcommand{\RNum}[1]{\uppercase\expandafter{\romannumeral #1\relax}}

\usepackage[dvipsnames]{xcolor} %

\newcommand{\nns}{LINNA}
\newcommand{\linna}{LINNA}

\newcommand{\emcee}{{\sc emcee}}
\newcommand{\cosmolike}{{\sc CosmoLike}}
\newcommand{\clustercomb}{4$\times$2pt+N}
\newcommand{\allcomb}{6$\times$2pt+N}
\newcommand{\ttt}{3$\times$2pt}

\newcommand{\Omegam}{\Omega_{{\rm m}}}

\title{\boldmath LINNA: Likelihood Inference Neural Network Accelerator}

\author[a,b,c,1]{Chun-Hao To \note{Corresponding author.}}
\author[d]{Eduardo Rozo}
\author[d,e]{Elisabeth Krause}
\author[f]{Hao-Yi Wu}
\author[g,h,i]{Risa H. Wechsler}
\author[c]{Andr\'es N. Salcedo}
\emailAdd{to.87@osu.edu}
\emailAdd{erozo@arizona.edu}
\emailAdd{krausee@arizona.edu}
\emailAdd{hywu@boisestate.edu}
\emailAdd{rwechsler@stanford.edu}
\emailAdd{ansalcedo@arizona.edu}

\affiliation[a]{ Center for Cosmology and AstroParticle Physics (CCAPP), Ohio State University, Columbus, OH 43210, USA}

\affiliation[b]{  Department of Physics, Ohio State University, Columbus, OH 43210, USA}
\affiliation[c]{  Department of Astronomy, Ohio State University, Columbus, OH 43210, USA}
\affiliation[d]{  Department of Physics, University of Arizona, Tucson, AZ 85721, USA}
\affiliation[e]{  Department of Astronomy/Steward Observatory, University of Arizona, 933 North Cherry Avenue, Tucson, AZ 85721-0065, USA}
\affiliation[f]{  Department of Physics, Boise State University, Boise, ID 83725, USA}
\affiliation[g]{  Department of Physics, Stanford University, 382 Via Pueblo Mall, Stanford, CA 94305, USA}
\affiliation[h]{  Kavli Institute for Particle Astrophysics \& Cosmology, P. O. Box 2450, Stanford University, Stanford, CA 94305, USA}
\affiliation[i]{  SLAC National Accelerator Laboratory, Menlo Park, CA 94025, USA
}

\abstract{
Bayesian posterior inference of modern multi-probe cosmological analyses incurs massive computational costs. %
For instance, depending on the combinations of probes, a
single posterior inference for the Dark Energy Survey (DES) data had a wall-clock time that ranged from 1 to 21 days using a state-of-the-art computing cluster with 100 cores. These %
computational costs have severe environmental impacts and the long wall-clock time slows scientific productivity. To address these difficulties, we introduce \nns{}: the Likelihood Inference Neural Network Accelerator.  Relative to the baseline DES analyses, \nns{} reduces the computational cost associated with posterior inference by a factor of 8--50. %
If applied to the first-year cosmological analysis of Rubin Observatory's Legacy Survey of Space and Time (LSST Y1), we conservatively estimate that \linna\ will save more than US $\$300,000$ on energy costs, while simultaneously reducing $\rm{CO}_2$ emission by $2,400$ tons. To accomplish these reductions, \nns{} automatically builds training data sets, creates neural network surrogate models, and produces a Markov chain that samples the posterior. We explicitly verify that \nns{}  
accurately reproduces the first-year DES (DES Y1) cosmological constraints derived from a variety of different data vectors with our default code settings, without needing to retune the algorithm every time.
Further, we find that \nns{} is sufficient for enabling accurate and efficient sampling for LSST Y10 multi-probe analyses.  We make \nns{} publicly available at ~\url{https://github.com/chto/linna}, to enable others to perform fast and accurate posterior inference in contemporary cosmological analyses. 
}
\begin{document}
\maketitle
\flushbottom

\section{Introduction}

Modern cosmological analyses typically employ a Bayesian framework, where posteriors are sampled using  Markov Chain Monte Carlo (MCMC) techniques.  Despite significant progress in the development of MCMC sampling algorithms \citep[e.g.][]{Bayesfast, karamanis2021zeus}, the computational requirements associated with performing these calculations can be significant. This is especially true for multi-probe analyses \cite{DESY1KP, KIDScombine,2018MNRAS.474.4894J,DESY3, Datapaper}, which typically require the introduction of many tens of nuisance parameters to account for systematic effects presented in the data. As a specific example, the joint analysis of clusters, weak lensing, and galaxy clustering of \cite*{Datapaper} required 26 nuisance parameters. To achieve a well-sampled posterior, the chains in that analysis required at least five million likelihood evaluations, totaling $\sim 50k$ CPU hours. That is, a single run had a wall-clock time of \textit{three weeks} using a state-of-the-art computing cluster with 100 cores. 

The primary difficulty in this type of analysis is that the theory model used to evaluate the likelihood is computationally expensive. One way to bypass this limitation is to rely on fast surrogates of said model \citep[e.g.][]{2008MNRAS.387.1575A, 2014MNRAS.439.2102A,2020MNRAS.499.5257P, 2022MNRAS.tmp...82S,2021arXiv211205889D}. However, the surrogates are typically built in pre-defined parameter spaces that are  narrower than those used in most cosmological analyses. That is, the use of surrogates that are constructed \textit{a priori} either limits the choice of priors that can be employed, or requires extrapolations of the surrogate outside the parameter regions over which the surrogate was calibrated. 

To address these shortcomings, as well as to avoid having to build surrogates of theory models on a case-by-case basis, we present a new posterior inference tool: the Likelihood Inference Neural Network Accelerator (\nns{}). \nns{} is designed to enable fast and accurate posterior inference with arbitrary priors through the use of automatically built surrogates of the theory model. Given the priors of the parameters, a theory model, the corresponding posterior function, and an efficient MCMC sampling algorithm, \nns{} automatically generates training data, trains a neural network emulator of the model predictions, and samples the posterior using the provided MCMC sampling algorithm. The training data for the neural network emulator is generated in an iterative fashion designed to achieve accurate posteriors. %

In this paper, we  demonstrate that \nns{} succeeds in accurately recovering the parameter posteriors derived from three different data vectors derived from the Dark Energy Survey \citep{DES} year 1 analyses (DES Y1):
\begin{enumerate}
    \item \ttt{}, a joint analysis of galaxy clustering, galaxy--galaxy lensing, and cosmic shear \cite{DESY1KP}.
    \item \clustercomb{}, a joint analysis of cluster--galaxy cross correlations, cluster lensing, cluster clustering, and cluster abundances \cite{Datapaper, Simpaper}.
    \item \allcomb{}, a joint analysis of data vectors in \ttt{} and \clustercomb{}. \cite{Datapaper}
\end{enumerate}   

Further, we explicitly demonstrate that \nns{} accurately reproduces the forecasted constraints for Rubin Observatory's Legacy Survey of Space and Time (LSST) Year 10 data set (see appendix \ref{app:LSST} for details).

This paper is organized as follows: In section \ref{sec:method}, we describe the design of \nns{}, the architecture of the neural network, the generation of training data, and the optimization of neural network parameters. In section \ref{sec:descusion}, we demonstrate that \nns{} can properly sample the posteriors of \allcomb{} (section \ref{sec6x2pt}), \clustercomb{} (section \ref{sec4x2pt}), and \ttt{} (section \ref{sec4x2pt}).  We further compare the performance of \nns{} to a promising fast sampler \textsc{Bayesfast}. The amount of reduction on $\rm{CO}_2$ emission by using \nns{} is estimated in section \ref{sec:co2}. A concluding remark is presented in section \ref{sec:conclu}. 

\section{Methods}
\label{sec:method}
\subsection{\nns{} design}
\begin{figure}
\centering
\includegraphics[width=0.7\textwidth]{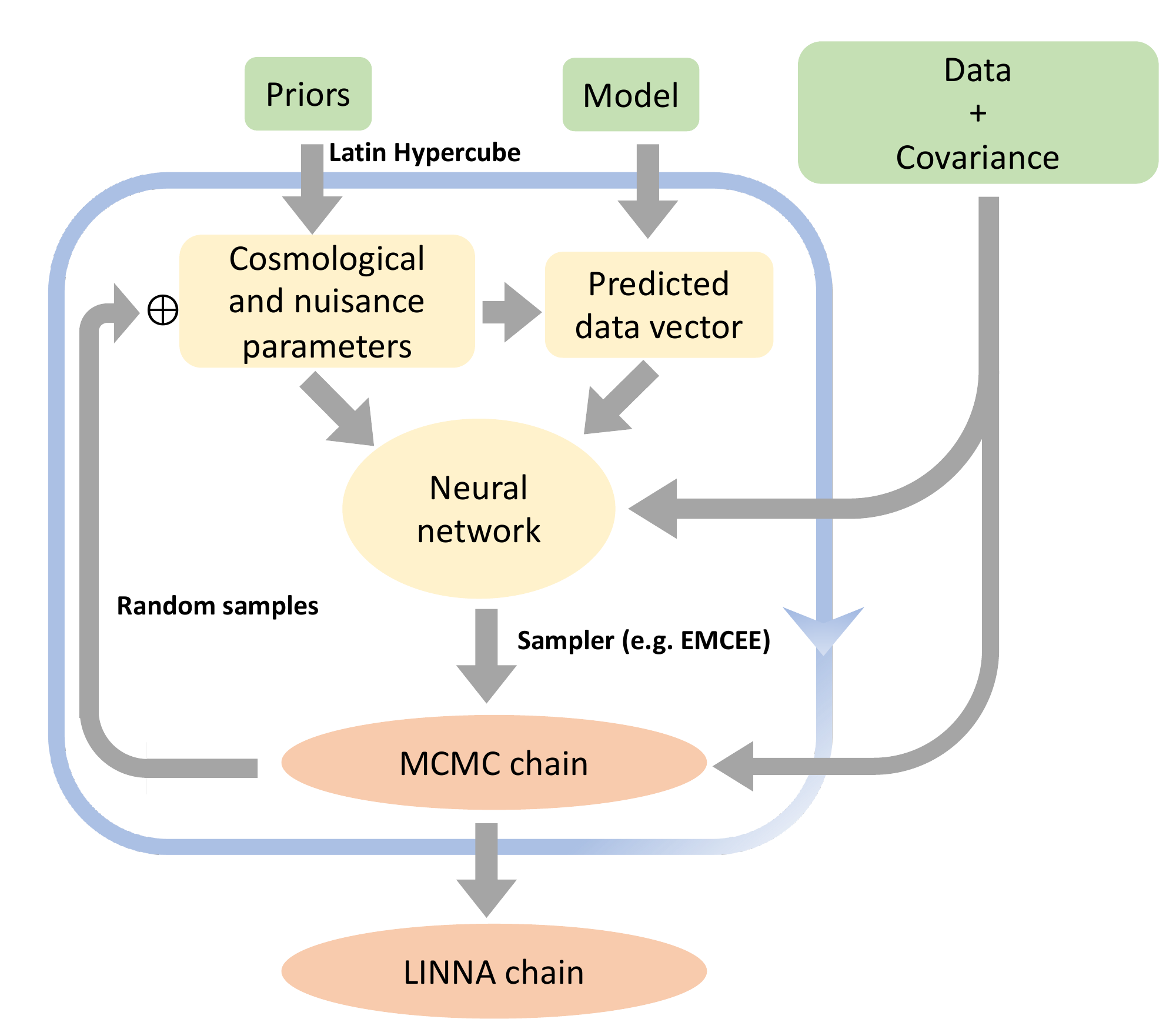}\hspace{-0.05\textwidth}
\caption{Outline of \nns{} structure. Green blocks indicate inputs provided by users and orange blocks represent outputs of \nns{}. Blocks within the blue rectangular are run iteratively. The symbol $\bigoplus$ represents the procedure of expanding the training data by randomly sampling the MCMC chain from the previous iteration. 
}
\label{fig:structure}
\end{figure}

LINNA uses neural networks to create fast surrogates of computationally expensive theory models.  This emulator can then be used in combination with
efficient posterior sampling algorithms to quickly recover accurate approximations of the parameter posteriors.
There are three main reasons why this approach is attractive:
\begin{enumerate}
    \item Within the context of posterior estimation, the parameter space of interest has a well-defined boundary: the prior. This boundary makes it possible to generate a training set that spans the entire parameter space. This ensures that the \linna\  emulator never extrapolates outside the domain over which it has been trained.
    \item The neural network surrogate is differentiable, allowing the use of more efficient sampling algorithms, such as the Hamiltonian Monte Carlo (HMC). In practice, existing HMC codes are not parallelized, which incurs a significant penalty in the wall-clock time required to generate an MCMC chain relative to less efficient but fully parallelized sampling schemes.%
    \item Because the posteriors are sampled using standard techniques, we may readily use well-tested convergence tests of the resulting chains.
\end{enumerate}

Despite these features, a significant problem remains.  In high-dimensional parameter spaces, the volume of reasonably defined priors can greatly exceed the volume of the posteriors. Consequently, uniform sampling of the prior is extremely inefficient: it is likely that only a few training points fall within one- to three- sigma region of the posteriors, where high accuracy of the neural network surrogates 
is most desirable. We solve this problem using an iterative approach. In particular, a crude neural network is trained on a training set that uniformly samples the priors. This neural network is then fed into a posterior sampling tool to produce a MCMC chain, from which additional training data is generated. In this way, we construct a training set that covers the entire prior volume and is also capable of accurately reconstructing the parameter posteriors with comparatively little training data.

In the first few iterations, the neural network surrogates are rough at best, so their high-confidence regions can be offset from those of the true posterior. To overcome this problem, for the first few iterations of the algorithm, we enlarge the parameter posteriors %
using %
a temperature parameter $T$ such that
\begin{equation}
    \chi^2_T = \chi^2_{\rm{NN}}/T^2,
\end{equation}
where $\chi^2_{\rm{NN}}$ is the $\chi^2$ value of the data calculated using neural network surrogates. We can think of this scaling as simply increasing all observational errors by a factor of T. The resulting broadening %
allows for the posteriors of the neural network surrogate model to be offset from the true posterior while still completely covering the same.  In this way, when we generate new training data from the surrogate posterior we are able to improve the performance of the network over the true parameter posterior, even if the latter was offset from that of the approximate model. %
 Initially, the temperature $T$ should be set to a large value to ensure that the training data spans a wide enough parameter space. In the subsequent iterations, $T$ can be gradually decreased due to the increasing accuracy of the neural network model over the region of interest. Empirically, we find that setting $T=4$ and $2$ for the first two iterations and $T=1$ for all subsequent iterations leads to good posteriors.

We summarize the workflow for \nns{} in figure~\ref{fig:structure}. First, we use Latin Hypercube sampling of the prior volume space to generate an initial set of training points. At each training point, we use the provided theory model to evaluate the expectation value of the observable. The resulting training points and model expectations are used to train a neural network to arrive at our first surrogate model. This surrogate model in combination with the provided likelihood function, which is scaled by the appropriate temperature parameter, is fed to a standard sampling algorithm to arrive at MCMC samples of the posterior. The resulting chain is then randomly sampled to expand the number of points in the parameter space at which the neural network is trained. The procedure is iterated, decreasing the temperature from iteration to iteration as described above.

We run \nns{} using the two publicly available posterior sampling code \emcee{} \citep{emcee} and \textsc{Zeus} \citep{karamanis2021zeus}. The convergence criteria are set based on the integrated autocorrelation time $\tau$ \citep{2010CAMCS...5...65G}. Specifically, let $\tau_i$ be the integrated autocorrelation time estimated using a chain of length $i$ for each walker. A chain is considered converged if the following three criteria are met: 
\begin{enumerate}
    \item The autocorrelation time estimation is stable, i.e., $\tau_i/\tau_{i-100}-1 < \eta$, where $\eta$ is a user-specified parameter.
    \item The number of steps per walker in units of the autocorrelation time is large, i.e., $i/\tau_i>\alpha$, where $\alpha$ is user-specified criterion.
    \item The mean and error of each parameter estimated from the MCMC chain are stable. We first select the last $\beta \times \tau_i$ samples for each walker to remove the burn-in steps. We compute the estimated mean of each parameter ($p$) using the first and second half of the selected MCMC samples ($\mu_{1,p}$ and $\mu_{2,p}$ respectively). The error of each parameter is also estimated using the standard deviation of the first and second half of the MCMC samples ($\sigma_{1,p}$ and $\sigma_{2,p}$ respectively). The chain is converged if $\rm{max}_p(\|\mu_{1,p}-\mu_{2,p}\|/\sigma_{2,p})<\delta_{\mu}$ and $\rm{max}_p(\|\sigma_{1,p}/\sigma_{2,p}-1\|)<\delta_{\sigma}$. 
\end{enumerate}
The above criteria are checked for every $100$ evaluation per walker.
The parameters $\eta$, $\alpha$, $\delta_{\mu}$, and $\delta_{\sigma}$ set the convergence criteria for the posterior sampling. We remove burn-in steps by keeping only the last $\beta\times \tau_i$ steps of each chain. Empirically, we have found the following settings provide accurate posteriors using 128 walkers: 
\begin{itemize}
    \item For \emcee{}, the values of the parameters $\eta$, $\alpha$,  and $\beta$ are set to $\eta=[0.03, 0.03, 0.02, 0.01]$, $\alpha=[5,5,10,15]$,  and $\beta=[2,2,5,4]$, where $[...]$ denotes the value for each iteration.  $\delta_{\mu}$ and $\delta_{\sigma}$ are set to $0.2$ and $0.15$ respectively for all iterations. %
    \item For \textsc{Zeus}, all parameters are kept the same, except for the value of $\alpha$ in the last  iteration. The speed of \textsc{Zeus} allows us to run a much longer chain with a modest increase of the overall run time. We therefore adopt $\alpha=50$ in the last iteration. 
    
\end{itemize}

\subsection{Neural network}\label{secnn}
\subsubsection{Neural network architecture}
\begin{figure}
\centering
\includegraphics[width=0.7\textwidth]{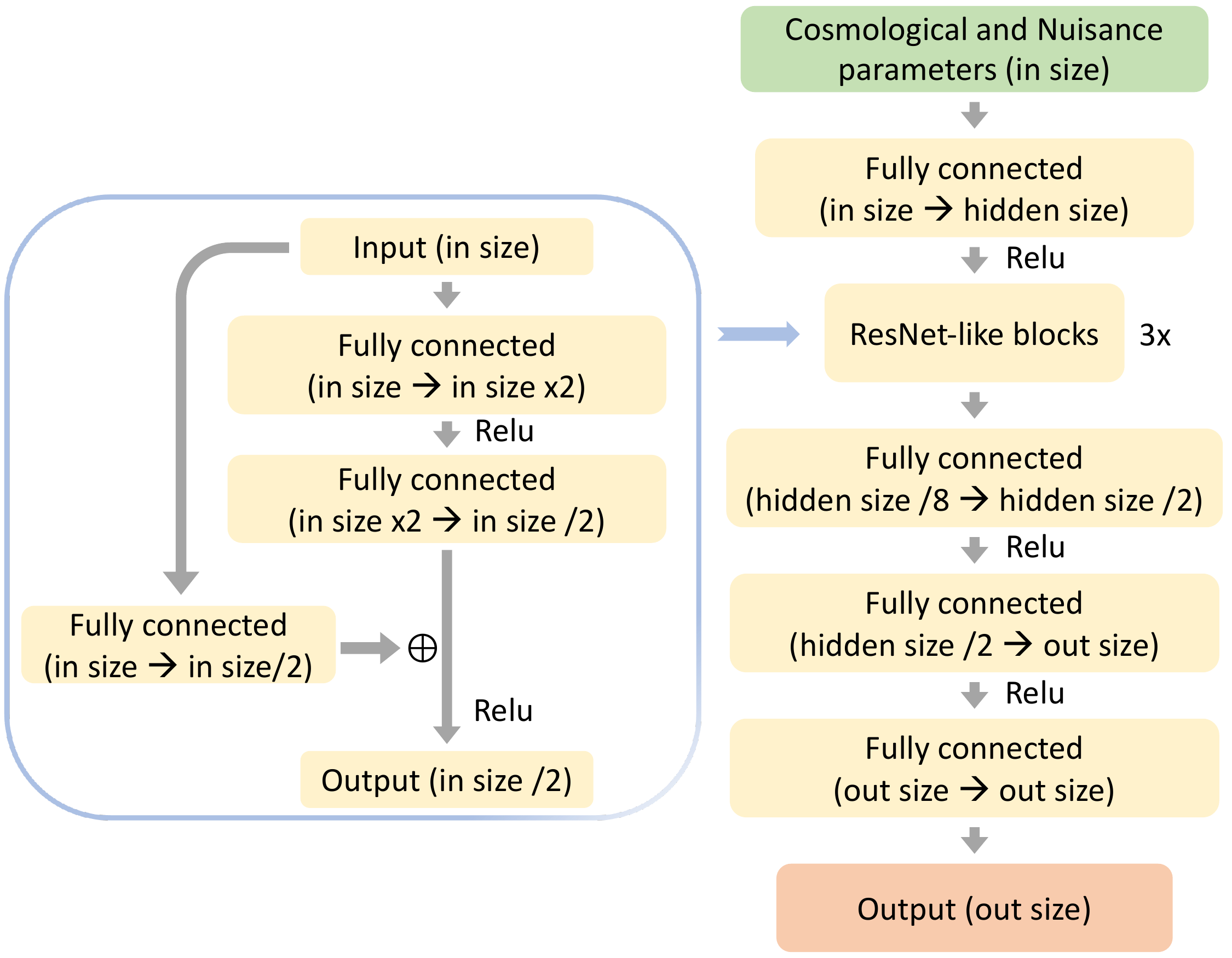}\hspace{-0.05\textwidth}
\caption{The architecture of the neural network in \nns{}. Green blocks represent input of the network and orange blocks represent the output. Numbers in paratheses show the size of input and output vectors of each block.  Hidden size is set to the minimum value of 32 times output vector size and $1000$. Relu represents the Rectified Linear Unit, a widely used nonlinear function. The symbol $\bigoplus$ denotes element-wise addition.} 
\label{fig:architecture}
\end{figure}

An artificial neural network (or simply neural network) is an interconnection of nodes that mimic the neural structures of the brain. The basic components of a neural network are nodes and connections between nodes. A node receives multiple inputs, calculates their weighted sums, adds an optional bias term, and passes the weighted sum through a non-linear function to produce the output. Connections between nodes specify the weights used in the weighted sum calculations. A neural network architecture specifies how nodes are connected, and it can drastically affect the performance of a neural network on a specific problem \cite[e.g.][]{DBLP:journals/corr/CanzianiPC16}. One commonly used neural network architecture simply arranges nodes into multiple groups (often called layers) and connects all the nodes from one group to another. This network structure is often referred to as a multilayer perceptron and has been widely used in the cosmology community \cite[e.g.][]{2007MNRAS.376L..11A, 2012MNRAS.424.1409A, 2018JCAP...10..028M, 2021arXiv211205889D}.

Despite their broad applicability, neural networks with multiple layers are difficult to train and their accuracy is not necessarily better than neural networks with fewer layers. This is known as the Degradation Problem: with increasing numbers of layers, the accuracy of a neural network gets saturated and can degrade \citep{resnet}. This problem can be mitigated via introductions of ``shortcut'' connections \citep{resnet}, which provide connections that skip one or more layers. Inspired by \citep{resnet}, we modify the widely used multilayer perceptron architectures, which consists of multiple fully connected layers, by adding three shortcut connections for \nns{}. The resulting neural network architecture is shown in figure~\ref{fig:architecture}.

\subsubsection{Loss function}
The loss function is judiciously chosen based on the task at hand, namely recovering accurate posteriors from the data. To that end, the loss function is motivated by two observations. 

First, the accuracy requirement of the surrogate model is set by the noise of the data, i.e., noisy data do not require accurate models. Consequently, instead of using a uniformly weighted loss function, we weight the differences between the training data and the surrogate model by the inverse of the covariance matrix of the data. 

Second, for the purposes of sampling posteriors, the surrogate model needs only to achieve high accuracy over high-likelihood regions. That is, whether a point in the parameter space has a posterior value that is $10^{-10}$ or $10^{-20}$ is irrelevant; we only need to know that this point is ruled out at high confidence. For this reason, we upweight the importance of those points in the parameter space that are a good fit to the data by dividing their contribution to the loss function by the $\chi^2$ of the model. Our loss function is therefore defined as 
\begin{equation}
    Loss = \left \langle \frac{(NN-M)^TC^{-1}(NN-M)}{(M-d)^TC^{-1}(M-d)} \right\rangle,
\end{equation}
where $NN$ denotes the output of the neural network, $M$ represents the (actual) model prediction, $C$ is the covariance matrix, $d$ is the data vector, and $\langle\dots\rangle$ refers to the mean over training points.

\subsubsection{Neural network optimization}
The accuracy of the neural network increases as the training set expands, at the cost of increasing the computation expense of generating training data and training the neural network.  We find that in each iteration of our algorithm, expanding the set of points used for training and validation by $10,000$ and $500$ respectively leads to a good compromise between accuracy and speed (see appendix  \ref{app:ntrain} for details). 

We also find that the performance improves if the training data are standardized before they are used to train the neural network. That is, the input of the neural network (i.e., the cosmological and nuisance parameters) is rescaled such that the entire training set has zero means and unit standard deviations. The model predictions are also rescaled judiciously. Specifically, let $\vec d_\alpha$ be the model prediction at the training point $\alpha$, and $d_\alpha^i$ be the $i^{{\rm th}}$ element of the data vector $\vec d_\alpha$.  We rescale each training point via 
\begin{equation}
    d_\alpha^i \rightarrow \frac{d_\alpha^i}{\sigma^i}
\end{equation}
where $(\sigma^i)^2$ is the $i^{{\rm th}}$ diagonal entry of the covariance matrix used to compute the likelihood function.  We further define
\begin{eqnarray}
\mu^i & = & {\rm median_\alpha}(d_\alpha^i) \\
s^i & = & {\rm MAD_{\alpha}}(d_\alpha^i),
\end{eqnarray}
where MAD refers to the median absolute deviation.  Notice that the median values that define $\mu^i$ and $s^i$ are over the set of training data (index $\alpha$), and are computed for each individual data point (index $i$).  With these quantities in hand, we further rescale the training data via
\begin{equation}
d_\alpha^i \rightarrow \frac{d_\alpha^i- \mu^i}{s^i}.
\end{equation}
We find that rescaling by the median/MAD performs significantly better than rescaling by the mean and standard deviation, as the latter two are much more sensitive to a few outliers in the training data. The original data vector $d^i$ can be recovered from the emulated data vector $\tilde d^i$ by applying the inverse transformation, i.e.
\begin{equation}
    d^i = (\tilde d^i s^i + \mu^i)\sigma^i.
\end{equation}

During %
training, the training set is randomly shuffled and split into batches, each containing $500$ points. The loss function is estimated using each batch of the training data and weights of the neural network are updated using the estimated loss function. We update the weights of the neural network using \textsc{adamw} \citep{adamw}, with an initial learning rate determined following methods described in \cite{smith2017cyclical} and an initial weight decay parameter of $10^{-4}$. 
 
The stop condition for the training of the neural network is set using the validation data. Specifically, after the network is updated using all batches of the training data, we evaluate the loss function on the validation data. 
The learning rate is decreased by half when the minimum loss in the validation data is not updated after $450$ steps. During the training, we use 200 steps to estimate the local derivative of the loss at a step. If the derivative is negative on the training set but positive on the validation set, the weight decay parameter is doubled to prevent over-fitting. The training process is automatically terminated when the minimum loss in the  validation data is not updated after $500$ steps.

The posterior is sampled using the trained neural network with the use of Intel Math Kernel Library for Deep Neural Networks \footnote{\url{https://github.com/rsdubtso/mkl-dnn}}, which increases the evaluation speed by more than a thousand times relative to not using it.   %

\section{Results and Discussion}
\label{sec:descusion}
We now test the performance of \nns{} in a real-world scenario. Specifically, we compare the posterior estimated using \nns{} to that estimated using \emcee{} to sample the posterior function implemented in \cosmolike{} \citep{cosmolike2016} (hereafter, the ``brute force method''). Consequently, for the purpose of this comparison, the results presented here are obtained using \emcee{} as the \nns{} sampler and \cosmolike{} as the \nns{} theory model. In practice, further improvements can be achieved using faster posterior sampling algorithms such as \textsc{Zeus} (see figure~\ref{fig:app:performance} for details). The comparisons described in the main body of the paper are mostly visual (see figures ~\ref{fig:allcon}, \ref{fig:allcon4x2pt}, and \ref{fig:allcon3x2pt}).  A more rigorous comparison of the \linna\ and ``brute force method'' chains can be found in Appendix~\ref{app:comparison_method}.

We perform all analyses using $128$ CPU cores and $1$ NVIDIA GeForce RTX 2080 Ti GPU on the Sherlock supercomputer\footnote{\url{https://www.sherlock.stanford.edu/}}. We consider three survey cosmology data vectors measured from first-year observations of DES (see a summary of these data vectors in figure 1 of  \cite{Datapaper}). Ranked by their computational cost, the three data sets are:
\begin{enumerate}
    \item \allcomb{} \citep*{Datapaper}: a joint analysis of all data vectors in \ttt{} and \clustercomb{}.  

\item \clustercomb{} \citep*{Datapaper,Simpaper}: a joint analysis of galaxy clustering, cluster--galaxy cross correlations, cluster clustering, cluster lensing, and cluster abundances.
    \item \ttt{} \citep{DESY1KP}: a joint analysis of galaxy clustering, galaxy--galaxy lensing, and cosmic shear. 
\end{enumerate}    
In these analyses, we sample 32 (28, 26) cosmological and nuisance parameters for \allcomb{} (\clustercomb{}, \ttt{}) with the same priors presented in \cite*{ Datapaper}.
\begin{figure*}
\centering
\includegraphics[width=1.0\textwidth]{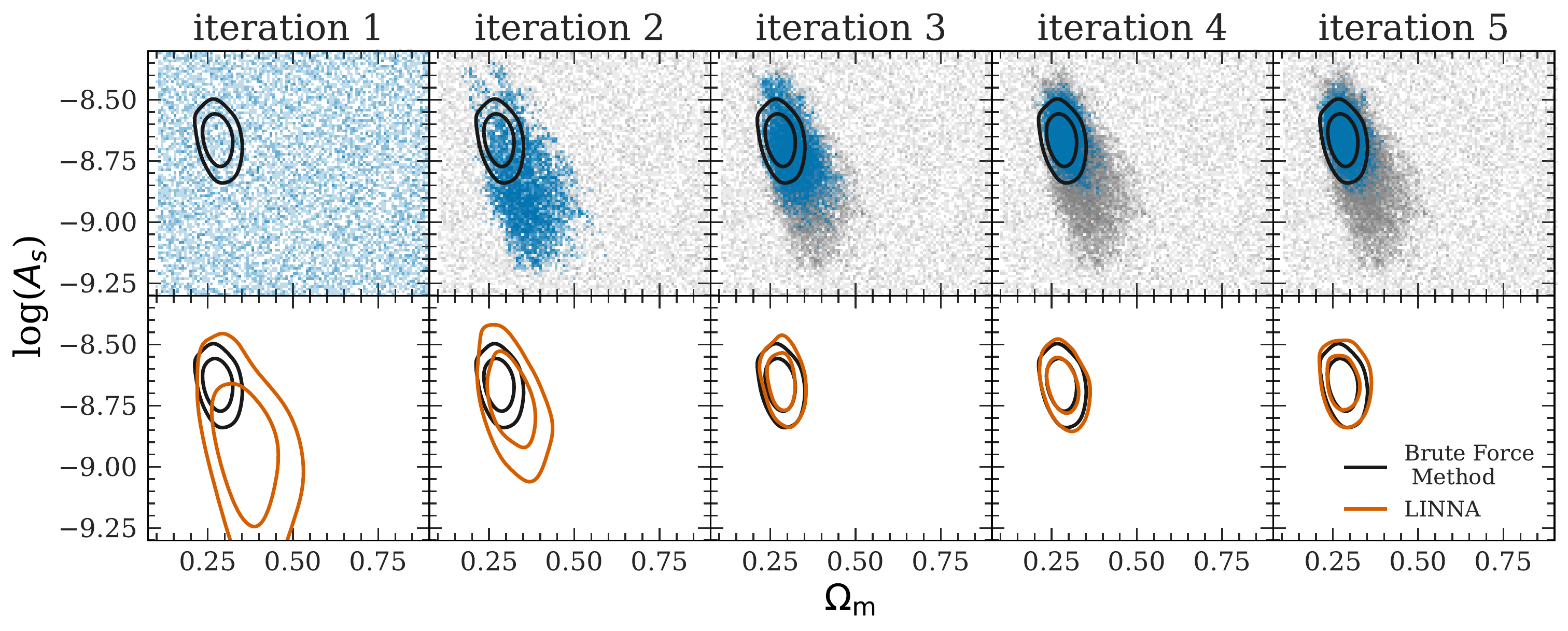}\hspace{-0.05\textwidth}
\caption{An illustration of the \nns{} algorithm in the $A_{\rm{s}}$--$\Omega_{\rm{m}}$ parameter subspace, when applied to the \allcomb{} data set. Top row: distributions of training data. Training data from previous iterations are shown as grey dots while newly added training data  are shown as blue dots. Black contours show the $68 \%$ and $95\%$ contours of the targeted posteriors as evaluated using \emcee{} to sample the posterior function implemented in \cosmolike{} (the brute force method).
Bottom rows: comparison of \nns{} posteriors (orange) to the target posterior (black) at each iteration. \nns{} shows strong consistency with the brute force method after three iterations. We note that while we only show $\Omega_{\rm{m}}$ and $A_{\rm{s}}$ in this plot, \nns{} shows strong consistency with the brute force method after three iterations in all other cosmological and nuisance parameters as well.}
\label{fig:iteration}
\end{figure*}
\subsection{\allcomb{} analysis}
\label{sec6x2pt}
figure~\ref{fig:iteration} shows the $68 \%$ and $95\%$  constraints on $A_{\rm{s}}$ and $\Omegam$ using \nns{} at each iteration.  
The blue points in the top row correspond to the $10,000$ newly added training data points at each iteration, while the grey points represent the cumulative training data from previous iterations. In the first iteration, the training set is generated by uniformly sampling the prior volume using a Latin Hypercube. %
For the purposes of Latin Hypercube sampling, the boundaries of the parameters with Gaussian priors are set to $\pm n \sigma$ around their mean values, where $n$ is a user-specified parameter. Throughout the analysis, we adopt $n=3$ because the Gaussian priors are only applied to nuisance parameters, whose priors are conservatively chosen. 
We see that \nns{} converges in just $3$ iterations, though we advocate running an additional high-precision iteration to verify the convergence. For this reason, we adopt four iterations as our standard \nns{} setting.

\begin{figure*}
\centering
\includegraphics[width=1.0\textwidth]{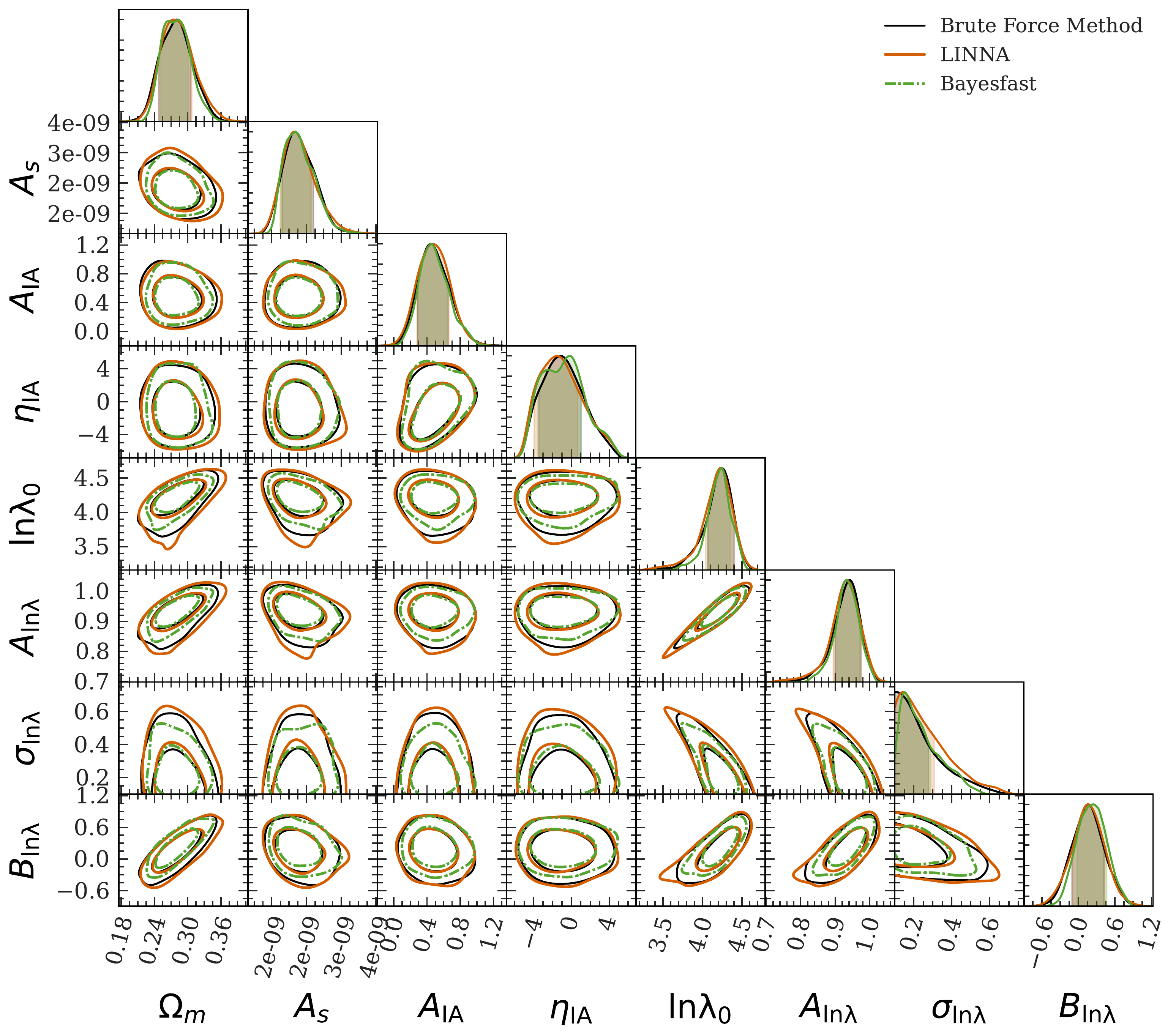}\hspace{-0.05\textwidth}
\caption{Constraints on important parameters when fitting to \allcomb{} after marginalizing over additional $24$ nuisance parameters. Black contours are obtained with the brute force method (\emcee{}+\cosmolike{}), orange contours are obtained with \nns{}, and green contours are obtained using \textsc{Bayesfast}. Contours show $68 \%$  and $95 \%$ constraints.}
\label{fig:allcon}
\end{figure*}

figure~\ref{fig:allcon} compares the final constraints obtained with \nns{} (orange contours) to those derived using the brute force method (black contours). The comparison is restricted to the most important parameters in the analyses: $\Omegam$, $A_{\rm{s}}$, two nonlinear intrinsic alignment parameters ($A_{\rm{IA}}, \eta_{\rm{IA}}$), and four richness--mass relation parameters ($\rm{ln}\lambda_{0}, A_{\rm{ln}\lambda}, B_{\rm{ln}\lambda}, \sigma_{\rm{ln}\lambda}$). We find excellent agreement between the two posteriors, with a shift in the mean parameter constraints of less than $0.2\sigma$. The one-sigma errors on individual parameters all agree to better than $12\%$. A more detailed comparison of the means and errors between the two posteriors is shown in figure~\ref{fig:app:multidimension_6x2pt}. Critically, \nns{} ran $\sim 10$ hours, compared to $\sim 3$ weeks for the brute force method.  A breakdown of total run time in each component of \nns{} is further shown in table \ref{apptable:break}. As noted earlier, the run time can be further reduced by decreasing the number of training data and by replacing \emcee{} with \textsc{Zeus} in \nns{}. Specifically, \nns{} can reach similar performances on all constrained cosmological and nuisance parameters with the run time down to $5.5$ hours. 

\begin{figure*}
\centering
\includegraphics[width=1.0\textwidth]{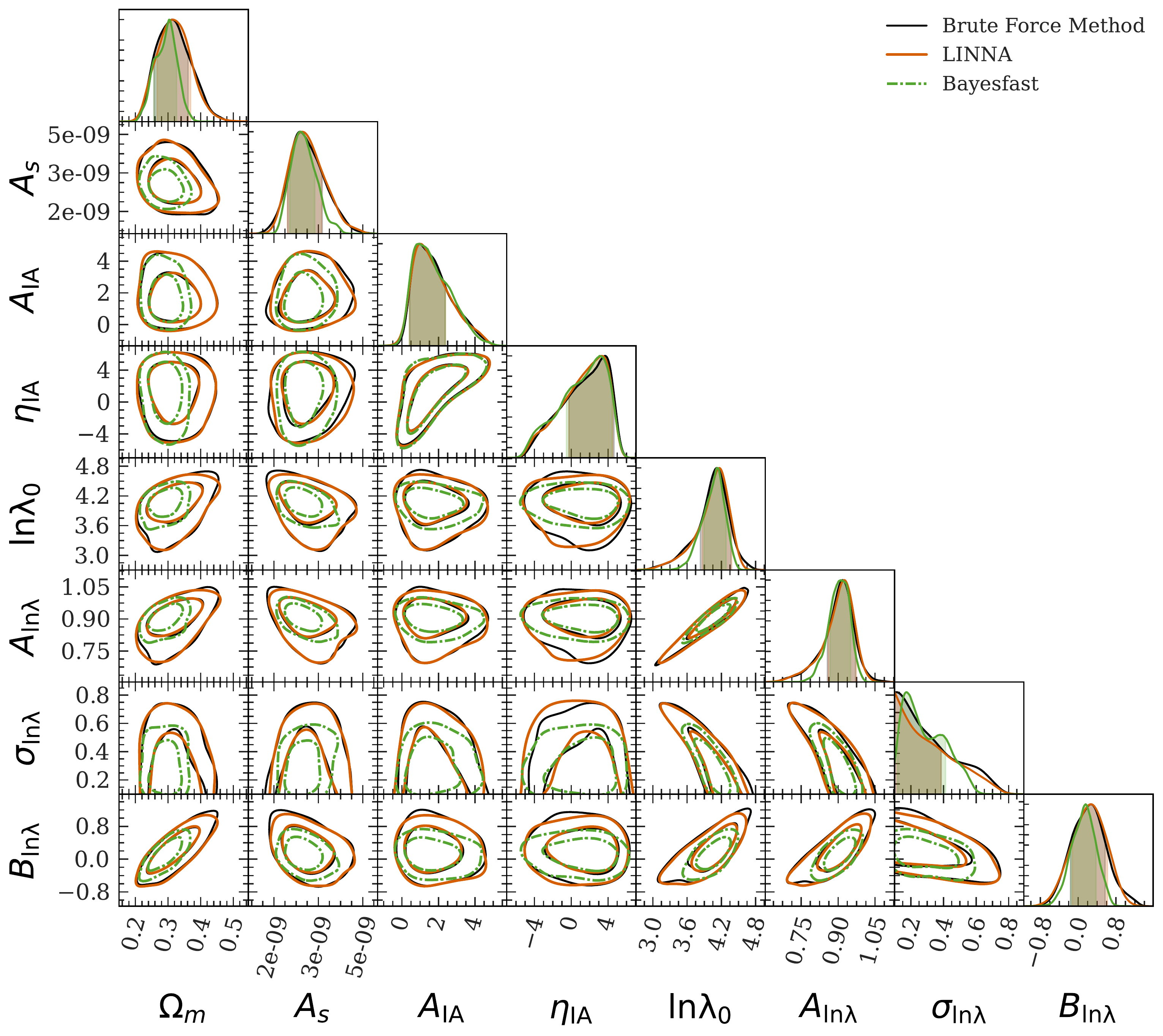}\hspace{-0.05\textwidth}
\caption{Similar to figure~\ref{fig:allcon} but on \clustercomb{} data vector.}
\label{fig:allcon4x2pt}
\end{figure*}

We also compare our results to those obtained using \textsc{Bayesfast}
\citep{Bayesfast} \footnote{\url{https://github.com/HerculesJack/bayesfast}}, another posterior inference acceleration code currently in development (green contours in figure~\ref{fig:allcon}). Similar to \nns{}, \textsc{Bayesfast} builds surrogate models by fitting quadratic polynomials to training data set in an iterative fashion. At each iteration, \textsc{Bayesfast} samples the approximate posterior using the No U-Turn Hamiltonian Monte Carlo algorithm \citep{NUTS}. After the last iteration, it importance samples the chain by comparing the likelihood evaluated using the surrogate model and the likelihood evaluated using the full theory model. %
Using the same MCMC convergence criteria and hardware as \nns{}, %
we find that \textsc{Bayesfast} converges in $5.8$ hours, a run time comparable to that of \nns{} using \textsc{Zeus} with $2000$ training points.
However, figure~\ref{fig:allcon} demonstrates that the accuracy in the recovered posteriors of \textsc{Bayesfast} is slightly worse than \nns{}.

\subsection{\clustercomb{} analysis and \ttt\ analysis}
\label{sec4x2pt}
\begin{figure}
\centering
\includegraphics[width=1.0\textwidth]{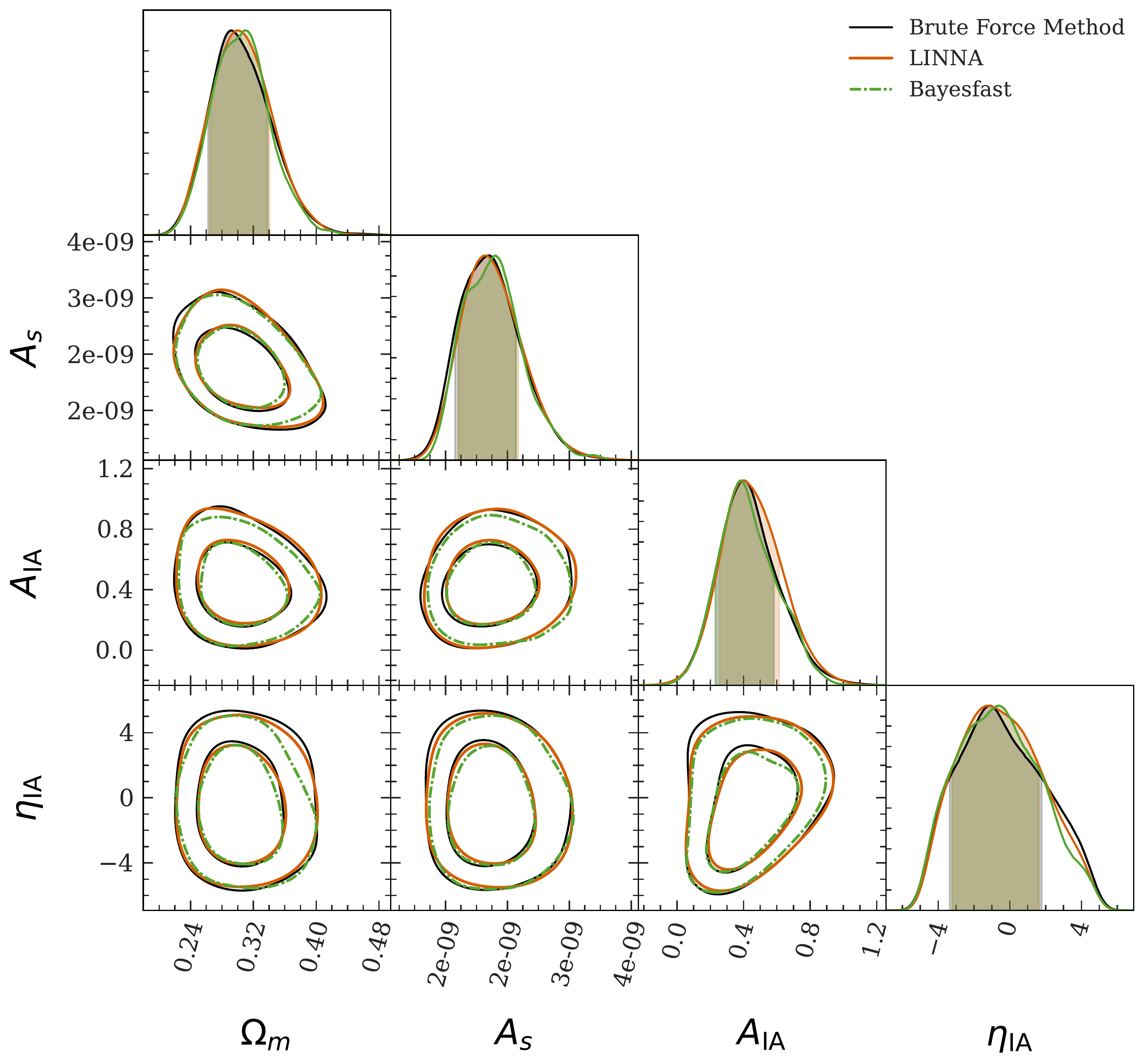}\hspace{-0.05\textwidth}
\caption{Similar to figure~\ref{fig:allcon} but on \ttt{} data vector.}
\label{fig:allcon3x2pt}
\end{figure}

figure~\ref{fig:allcon4x2pt} compares the posteriors derived for the \clustercomb{} data vector using \nns{} (orange contours), the brute force method (\emcee{}+\cosmolike{}, black contours) and \textsc{Bayesfast} (green contours). As before, we restrict the figure to the most important parameters for the \clustercomb{} data vector. All samplers are run with exactly the same setting as for the \allcomb{} analysis. The agreement between \nns{} and the brute force method is remarkable and demonstrates that \nns{}'s hyperparameters do not need to be retuned when running in subsets of the original data used to tune \nns{}. This is not a trivial statement:  \textsc{Bayesfast} posteriors, which were only modestly biased for \allcomb{} data set, are now much more strongly biased than before. 
figure~\ref{fig:allcon3x2pt} compares the posteriors derived for the \ttt{} data vector from \nns{} (orange contours), the brute force method (\emcee{}+\cosmolike{}, black contours), and \textsc{Bayesfast} (green contours).
Here, \nns{} and \textsc{Bayesfast} are both in excellent agreement  with the brute force method. 

As was the case for \allcomb{}, using \nns{} for these analyses leads to significant reductions in run time relative to the brute force method (\emcee{}+\cosmolike{}). With the baseline setting, \nns{} takes 11.5 (5.4) hours to sample the \clustercomb{} (\ttt{}) posteriors, while the brute force method takes $\sim 4$ (1) days respectively.  Again, \nns{} can be further sped up while maintaining the same accuracy on all constrained parameters by decreasing the number of training data and by using faster samplers. With the same fast setting as used for \allcomb{}, the run time of \nns{} is $4.2$ and $3.5$ hours for  \clustercomb{} and \ttt{} respectively. 

\subsection{Environmental Impact}
\label{sec:co2}
Supercomputers have been recognized as a significant contributor to the $\rm{CO}_2$ footprint of astronomical research \citep{2020NatAs...4..843S,2021NatAs...5..857B,2021NatAs...5.1195V}. To estimate the potential reductions of environmental impact by using \nns{}, we first calculate the $\rm{CO}_2$ emission savings associated with running one posterior inference analysis using \nns{} versus the current standard approach (e.g. \emcee{}+\cosmolike{}). \cite{2020NatAs...4..843S} estimates that the energy consumption per CPU core is $\sim 53W$. The maximum electricity consumption of a NVIDIA GeForce RTX 2080 Ti GPU on the Sherlock supercomputer is $280$W. Assuming the GPU is run with the maximum power consumption during the entire analysis, we find that \nns{} saves 12 billion Joule of electricity consumption per posterior inference analysis for \allcomb{} data vector. Assuming an average energy cost of US $\$0.10$ per kW-hour, as appropriate for the United States, this corresponds to US $\$330$ per posterior inference analysis. We translate this energy consumption to the amount of $\rm{CO}_2$ emission using the EPA Carbon Offset Calculator \footnote{\url{https://www.epa.gov/energy/greenhouse-gas-equivalencies-calculator}} and find a reduction of 2.4 tons $\rm{CO}_2$ emission per analysis.

Modern cosmology analyses typically require hundreds of posterior inference analyses to validate the analysis choices and modeling pipelines. For example, the DES collaboration performed over 900 posterior inference analyses to analyze the \ttt{} data set measured from the first three years observations of DES. %
Therefore, we estimate that the publications of the cosmological constraints derived from first-year observations of Rubin observatory's Legacy Survey of Space and Time (LSST Y1) will require $\sim 1,000$ posterior inference analyses. Assuming each posterior inference analysis of LSST Y1 has similar computational costs to those of the \allcomb{} data set of DES Y1,  we estimate that adopting \nns{} for the LSST Y1 analysis will save upwards of US $\$300,000$ in energy costs, while simultaneously reducing the $\rm{CO}_2$ footprint of the analysis by $2,400$ tons of $\rm{CO}_2$. This is roughly equivalent to $2,400$ transatlantic flights, or the yearly carbon footprint of $65$ astronomers \citep{2022arXiv220108748K}. In practice, we expect the number of chains is likely to exceed our estimate, as we anticipate the data will be analyzed using many different theoretical approaches.  In other words, relative to the DESY3 analyses, we have not scaled our estimates with collaboration size or model complexity in any way.

\section{Conclusions}
\label{sec:conclu}

Posterior inference for modern survey cosmological analyses is computationally expensive. This is mostly due to the increasing complexity of the theory models needed to yield unbiased cosmological constraints from unprecedentedly precise measurements. This situation becomes more challenging when one considers multi-probe analyses, in which one analyzes different cosmological probes simultaneously to yield more robust and precise cosmological constraints. The complex data vector of multi-probe analyses requires one to sample more nuisance parameters that describe systematics of each individual probe, thereby severely increasing the dimensionality of the posteriors. %
These difficulties are especially pronounced in the case of cluster cosmology.
  Modeling of cluster-related two-point correlation functions %
requires marginalization over observable--mass relation and cluster-related systematics, such as selection biases \citep*[e.g.][]{Simpaper}, cluster mis-centering \citep[e.g.][]{Tomclusterlensing}, and correlation coefficients between different cluster mass--observable connections \citep[e.g.][]{SPTClusters2019}. These additional modeling complexities %
  significantly increase the computational cost of posterior inference. %
For instance, %
the run time for the \ttt\  analysis of the first-year data from the Dark Energy Survey
  \citep{DESY1KP} %
  was $\sim$ 1 day.
With the same hardware, the corresponding \allcomb{} analysis \citep*{Datapaper} took $\sim 3$ weeks.

In this work, we expedite the posterior inference by constructing a surrogate of expensive cosmological models using a deep neural network (see figure~\ref{fig:architecture}). We design an iterative process to build a training data set that both spans the entire prior volume and reasonably samples the posterior with a limited amount of total training data (summarized in figure~\ref{fig:structure}). The trained surrogate model can be combined with standard posterior sampling tools to perform accurate Bayesian posterior inference analyses. We test the developed posterior inference accelerator \nns{} on data vectors measured from the Dark Energy Survey first-year observations. In particular, we consider three data vectors: \ttt{} (a joint analysis of galaxy clustering, galaxy--galaxy lensing, and cosmic shear), \clustercomb{} (a joint analysis of cluster--galaxy cross correlations, cluster lensing, cluster clustering, and cluster abundances), and \allcomb{} (a joint analysis of data vectors in \ttt{} and \clustercomb{}). The results of our tests can be summarized as follows: 
\begin{enumerate}

    \item In all data vectors, the posteriors obtained from \nns{} agree with those from the brute force method that  uses \emcee{} to sample the posterior implemented in \cosmolike{}. The hyperparameters of \nns{} are fine-tuned for \allcomb{}, but stay the same for \clustercomb{} and \ttt{}. %
    \item In the \allcomb{} (\clustercomb{}, \ttt{}) analysis, the brute force method took $\sim 3$ weeks ($\sim 4$ days, $\sim 1$ day) to sample the posterior using 128 CPUs while \nns{} took 10.1 (11.5, 5.4) hours using 128 CPUs and $1$  NVIDIA GeForce RTX 2080 Ti GPU. With the same performance on all constrained parameters, the run time of \nns{} can be further reduced to $5.5$ ($4.2, 3.5$) hours by reducing the number of training data and adopting \textsc{Zeus} as the posterior sampler. 
    \item %
    The reduced computational demands lead to a reduction of 12 billion Joules of electricity consumption per posterior inference analysis. We estimate that using \nns{} for the LSST Y1 cosmological analysis will save $\$300,000$ in energy costs, while simultaneously reducing $\rm{CO}_2$ emission by $2,400$ tons, equivalent to the annual carbon footprint of $\sim 65$ astronomers. 
\end{enumerate}

While we have demonstrated the performance of \nns{} using DES Y1 data set, we have also explicitly verified that \nns{} accurately reproduces the forecasted constraints for LSST Y10 for the \ttt{} and \allcomb{} data sets (see appendix \ref{app:LSST} for details). Further, although the run times measured in this paper are based on \cosmolike{} and \emcee{}, \nns{} allows users to replace these components with other modeling codes and different samplers such as \textsc{Zeus} \citep{karamanis2021zeus} or \textsc{MultiNest} \citep{Multinest} to further speed up the analysis. 
We make the \nns{} package publicly available at \url{https://github.com/chto/linna}, in the hope that it will prove useful for ongoing and future astronomical and cosmological analyses.

For discussions on an independently developed likelihood acceleration tool for LSST Year 1 \ttt{} analyses, we refer the reader to  \cite{Supranta}.

\section*{Acknowledgements}
CHT thanks Wei-Lin Chiang for helpful discussions on GPUs and neural networks and David Weinberg for insightful discussions. This work received support from the United States Department of Energy, Office of High Energy Physics under Award Number DE-SC-0011726 and 
under contract number DE-AC02-76SF00515 to SLAC National Accelerator Laboratory. ER is supported by DOE grant DE-SC0009913, NSF grant 2009401, and a Cottrell Scholar award. HYW is supported by DOE Grant DE-SC0021916 and NASA grant 15-WFIRST15-0008.

We acknowledge the use of GetDist \citep{Lewis:2019xzd}, Pytorch \citep{paszke2019pytorch}, Matplotlib \citep{Hunter:2007}, and NumPy \citep{numpy} for the analyses.
Some of the computing for this project was performed on the Sherlock cluster at Stanford. We would like to thank Stanford University, the Kavli Institute for Particle Astrophysics and Cosmology, and the Stanford Research Computing Center for providing computational resources and support that contributed to these research results.

\bibliography{sample.bib} %

\appendix
\section{Multidimensional comparison of two MCMC chains}
\label{app:comparison_method}

\begin{figure*}
\centering
\includegraphics[width=1.0\textwidth]{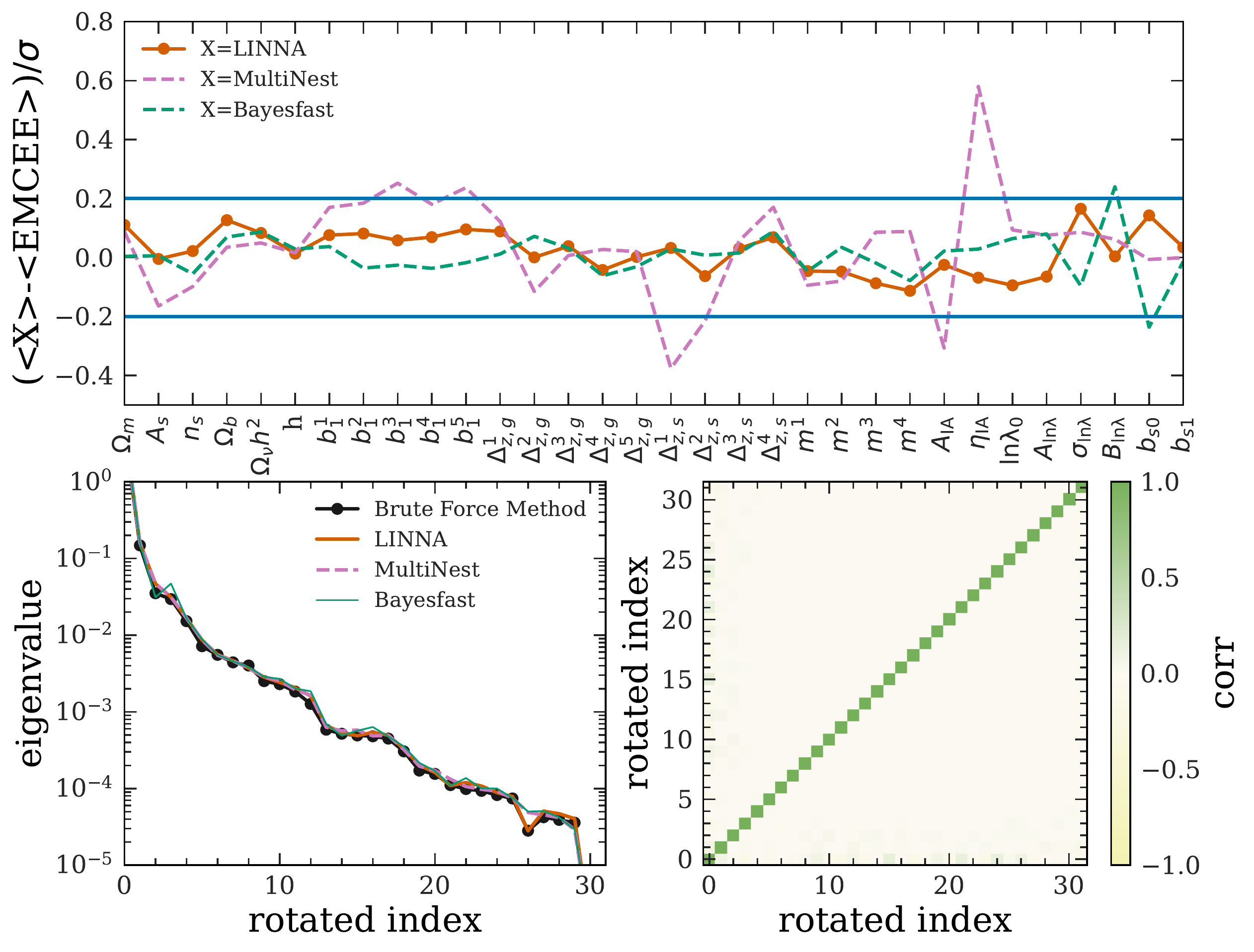}\hspace{-0.05\textwidth}
\caption{A comparison of multidimensional posteriors obtained with \nns{} and with \emcee{} from the \allcomb{} data vector. Top: a comparison of parameter means obtained with different samplers to $1\sigma$ uncertainties using \emcee{} chains. Bottom left: a comparison of eigen-value of parameter covariance matrix estimated using \emcee{} chains (black) and diagonal value of parameter covariance matrix estimated using \nns{} (
\textsc{Bayesfast}) chains after they are rotated to the eigen-space of former (orange (green)). Bottom right: the correlation matrix of parameters estimated using \nns{} chains after they are rotated to the eigen-space of parameter covariance matrix estimated using \emcee{} chains.}
\label{fig:app:multidimension_6x2pt}
\end{figure*}

We wish to compare \linna\ chains to those obtained using the brute force method.  To do so, we compare the means and errors for each of the parameters in the model, as shown in figure~\ref{fig:app:multidimension_6x2pt}.
In the top row of figure~\ref{fig:app:multidimension_6x2pt}, we 
show the difference in the mean between \linna\ and the brute force method for each model parameter, in units of the standard deviation of the same (as evaluated using the brute force method). We find that %
the difference between the two means is always less than $0.2\sigma$, with the median difference being $0.2\sigma$. The figure also compares the \emcee{} chains from the brute force method to those generated using \textsc{MultiNest}.  Notice the \linna\ posteriors are in better agreement with the brute force method than those estimated using \textsc{MultiNest}. %

\begin{figure*}
\centering
\includegraphics[width=1.0\textwidth]{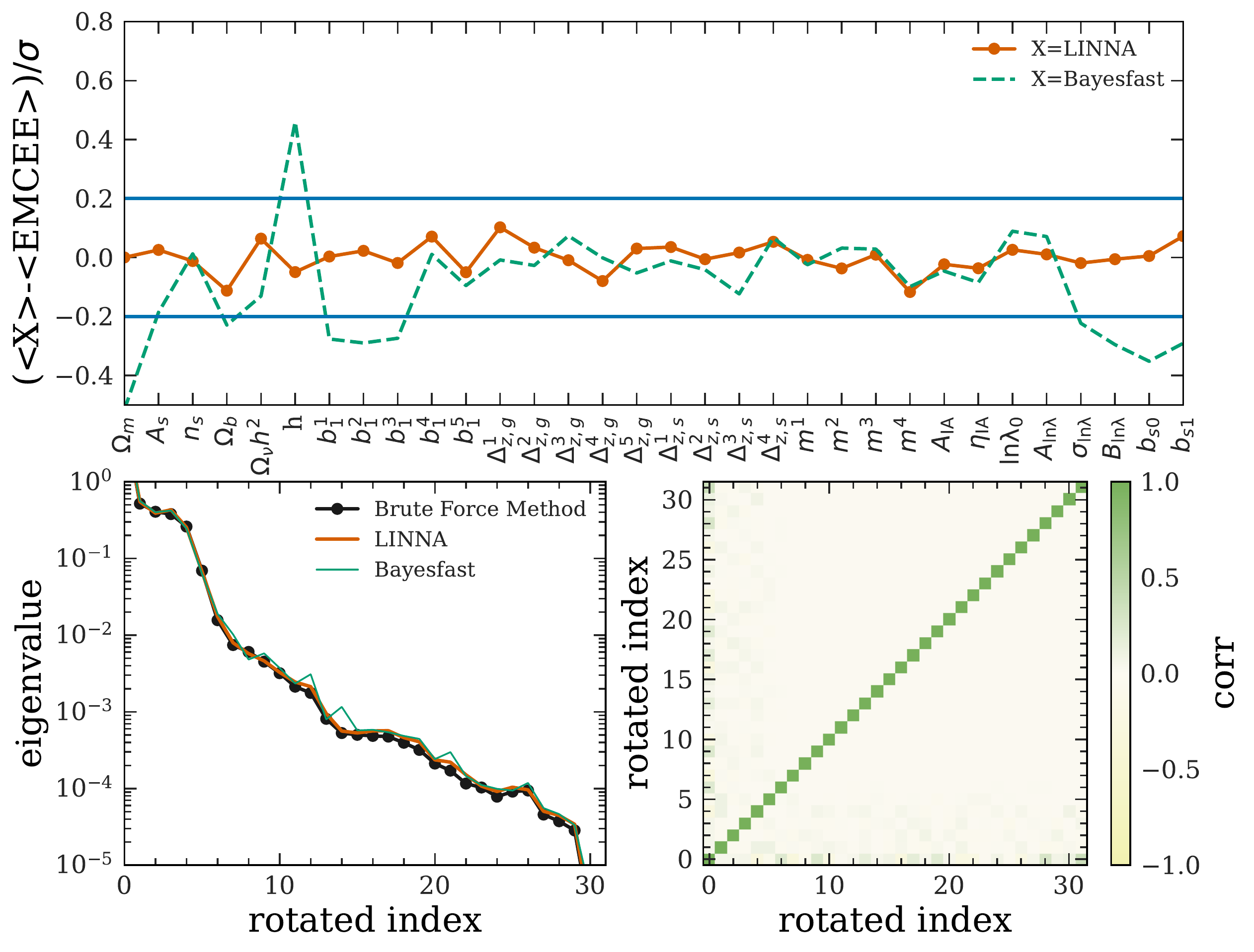}\hspace{-0.05\textwidth}
\caption{Similar to figure~\ref{fig:app:multidimension_6x2pt} but on the \clustercomb{} data vector.}
\label{fig:app:multidimension_4x2pt}
\end{figure*}

\begin{figure*}
\centering
\includegraphics[width=1.0\textwidth]{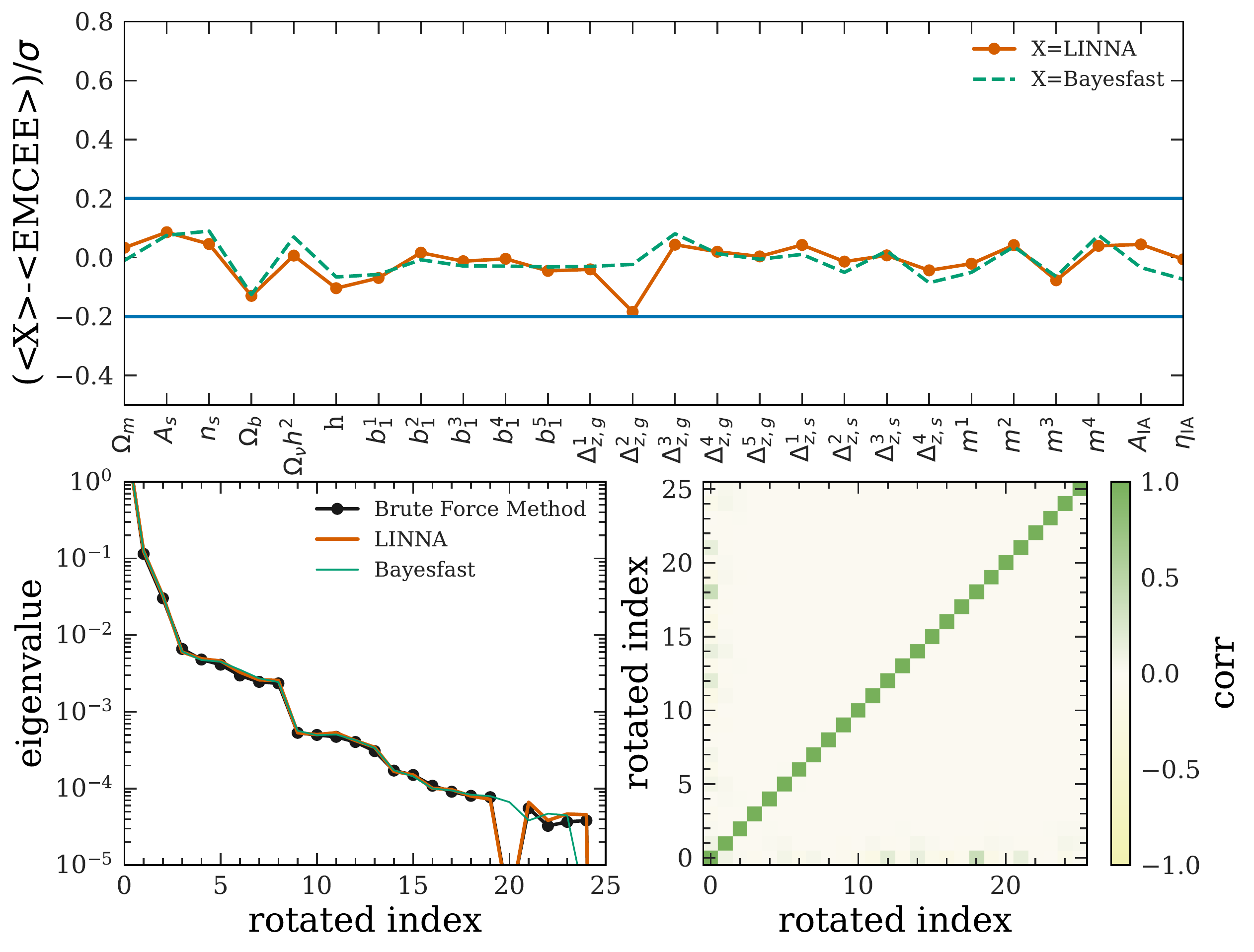}\hspace{-0.05\textwidth}
\caption{Similar to figure~\ref{fig:app:multidimension_6x2pt} but on the \ttt{} data vector.}
\label{fig:app:multidimension_3x2pt}
\end{figure*}

We further compare the uncertainty estimations from \nns{} and those from the brute force method in the bottom row of figure~\ref{fig:app:multidimension_6x2pt}. To make this comparison, we compute the parameter covariance matrix using chains from \nns{} and chains from the brute force method. We rotate the covariance matrix estimated using the \nns{} chain to the eigen-space of the covariance matrix estimated using the chain from the brute force method (reference covariance matrix hereafter). If the two covariance matrices are consistent, the rotated covariance matrix will be mostly diagonal with the diagonal value being identical to the eigen-value of the reference covariance matrix. In the bottom two panels of figure~\ref{fig:app:multidimension_6x2pt}, we show that the rotated covariance matrix is mostly diagonal and the diagonal value is nearly identical to  the eigen-value of the reference covariance matrix.
Similar comparisons for \clustercomb{} and \ttt{} data vectors are shown in figure~\ref{fig:app:multidimension_4x2pt} and \ref{fig:app:multidimension_3x2pt}.

\section{Optimizations of the number of training data and run-time breakdowns}
\label{app:ntrain}
\begin{figure}
\centering
\includegraphics[width=1.0\textwidth]{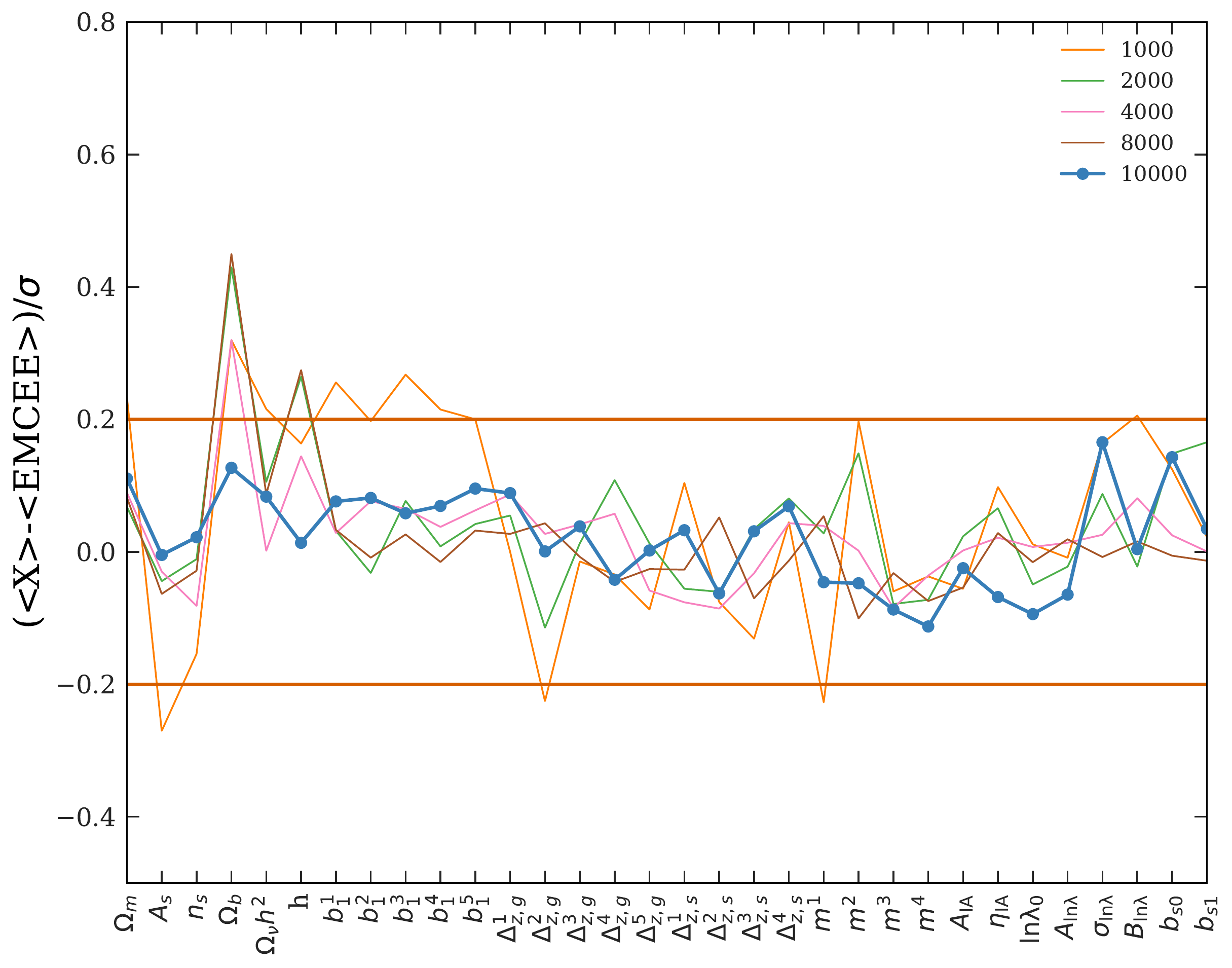}\hspace{-0.05\textwidth}
\caption{A comparison of \nns{} performance with different numbers of training data points. The $y$ axis shows the shifts in parameter mean estimations using \nns{} and using the brute force method in the unit of $1\sigma$ uncertainties. Different color lines correspond to \nns{} having $1000-10000$ additional training data points per iteration.}
\label{fig:app:performance}
\end{figure}

In figure~\ref{fig:app:performance}, we compare the performance of \nns{} with different numbers of training data points added in each iteration. We find that as long as the number of training data points added in each iteration is more than 2000, \nns{} leads to less than $0.2\sigma$ shifts in all parameters that are meaningfully constrained by the data. With 10000 training data points added in each iteration, \it all \rm parameters are shifted by less than $0.2 \sigma$ when compared to the brute force method. We thus 
adopt adding $10000$ training data points in each iteration as the default option for \nns{}, but note that $2000$ training data points could lead to a reasonable performance in all constrained parameters. In table \ref{apptable:break}, we show the breakdown of the total run time of \nns{} with $10000$ and $2000$ training data points added in each iteration respectively. %
The total run time reduces from 10.1 hours to $6.89$ hours when reducing the number of training data points from $10000$ to $2000$. We further note that in the case of $2000$ training data points, the \nns{} run time is dominated by the MCMC sampling. One could further speed up \nns{} using faster samplers such as \textsc{Zeus}. We find that using \textsc{Zeus} reduces the run time for the $2000$ training data point case to $5.52$ hours while maintaining the same performance.

\begin{table}
\caption{Breakdowns of \nns{} run time in each component using the default option. Model shows the amount of time spent in generating the training data using \cosmolike{}. Train shows the time spent in training the neural network. MCMC shows the time spent in sampling the posterior using \emcee{}. Numbers in each panel are in units of hours.
Numbers in the parentheses show run time for \nns{} with 2000 training data points in each iteration.
}
\label{apptable:break}
\centering
\begin{tabular}{lrrrrr}
\multicolumn{3}{l}{{\large Analysis: \allcomb{}}}&\\
\hline
\hline
& iteration 1   & iteration 2   & iteration 3   & iteration 4   & Total        \\
\hline
 Model & 1.12 (0.46)   & 0.67 (0.18)   & 0.74 (0.19)   & 0.78 (0.19)   & 3.31 (1.02)  \\
 Train & 0.51 (0.56)   & 0.61 (0.61)   & 1.23 (0.56)   & 1.25 (0.69)   & 3.60 (2.42)  \\
 MCMC  & 0.29 (0.30)   & 0.28 (0.25)   & 0.51 (0.58)   & 2.08 (2.33)   & 3.16 (3.46)  \\
 Total & 1.93 (1.32)   & 1.55 (1.05)   & 2.48 (1.33)   & 4.11 (3.21)   & 10.07 (6.89) \\
\hline
\hline\\
\multicolumn{3}{l}{{\large Analysis: \clustercomb{}}}&\\
\hline
\hline
       & iteration 1   & iteration 2   & iteration 3   & iteration 4   & Total         \\
\hline
 Model & 1.05 (0.74)   & 0.69 (0.25)   & 0.71 (0.26)   & 0.72 (0.26)   & 3.18 (1.51)   \\
 Train & 0.73 (0.22)   & 0.89 (0.38)   & 1.09 (0.52)   & 1.57 (0.52)   & 4.28 (1.64)   \\
 MCMC  & 0.26 (0.44)   & 0.30 (0.48)   & 0.61 (1.57)   & 2.82 (6.76)   & 3.99 (9.25)   \\
 Total & 2.05 (1.40)   & 1.88 (1.11)   & 2.42 (2.35)   & 5.11 (7.54)   & 11.45 (12.40) \\
\hline
\hline\\
\multicolumn{3}{l}{{\large Analysis: \ttt{}}}&\\
\hline
\hline
       & iteration 1   & iteration 2   & iteration 3   & iteration 4   & Total       \\
\hline
 Model & 0.55 (0.35)   & 0.14 (0.04)   & 0.14 (0.08)   & 0.14 (0.04)   & 0.99 (0.51) \\
 Train & 0.48 (0.26)   & 0.63 (0.62)   & 0.80 (0.33)   & 1.25 (0.46)   & 3.16 (1.67) \\
 MCMC  & 0.20 (0.21)   & 0.21 (0.20)   & 0.29 (0.28)   & 0.56 (0.57)   & 1.25 (1.26) \\
 Total & 1.23 (0.82)   & 0.98 (0.86)   & 1.24 (0.70)   & 1.95 (1.06)   & 5.39 (3.44) \\
\hline
\hline

\end{tabular}
\end{table}

\section{LSST Year 10 forecast}
\label{app:LSST}

We evaluate the expected performance of \nns{} on %
ten-year LSST (LSST Y10) data
using simulated data vectors. We construct the data vector using \cosmolike{}, using parameters described in \cite{LSST}. For details, we refer the reader to To, Krause et al. in prep. Here, we provide a short description. The lens galaxy samples are split into 10 equally spaced photometric redshift bins from $z=0.2$ to 1.2. The number density is $48$ $\rm{arcmin}^{-2}$. For the source galaxy samples, we assume the number density to be $27$ $\rm{arcmin}^{-2}$ and split them into five redshift bins from $z=0.2$ to $4$. The cluster samples are generated using the richness--mass relation obtained in \cite{Datapaper}, and are split into four equally spaced redshift bins from $z=0.2$ to $1$. In the analysis, we assume a wCDM cosmological model with $\rm{w}_0$ and $\rm{w}_a$ describing the dark energy equation of state and its redshift dependence. 

\begin{figure*}
\centering
\includegraphics[width=1.0\textwidth]{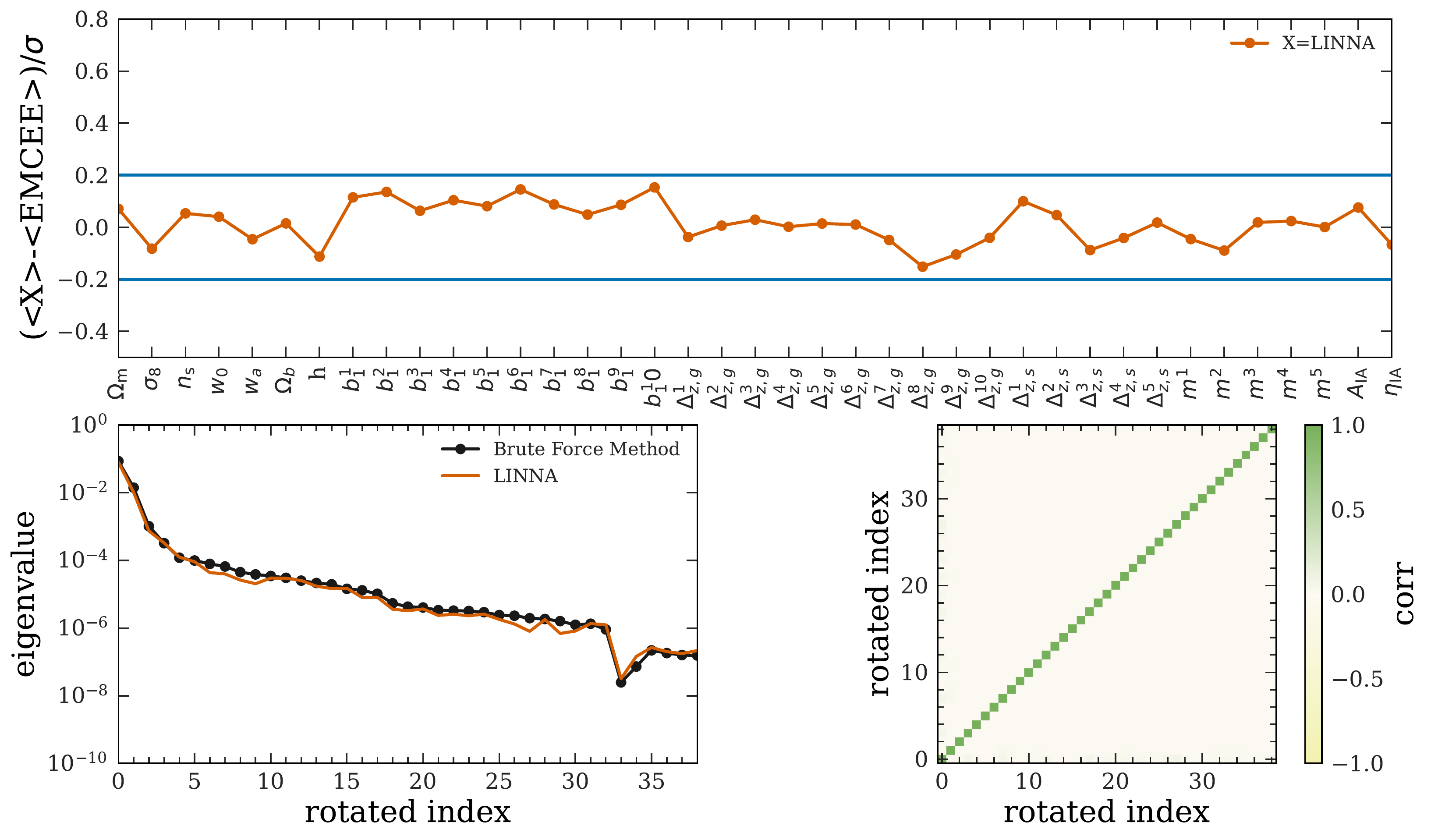}\hspace{-0.05\textwidth}
\caption{Performance of \nns{} on the \ttt{} analysis using the simulated LSST Y10 data vector. Details of this plot are similar to figure~\ref{fig:app:multidimension_6x2pt}.}
\label{app:LSST3x2pt}
\end{figure*}

\begin{figure*}
\centering
\includegraphics[width=1.0\textwidth]{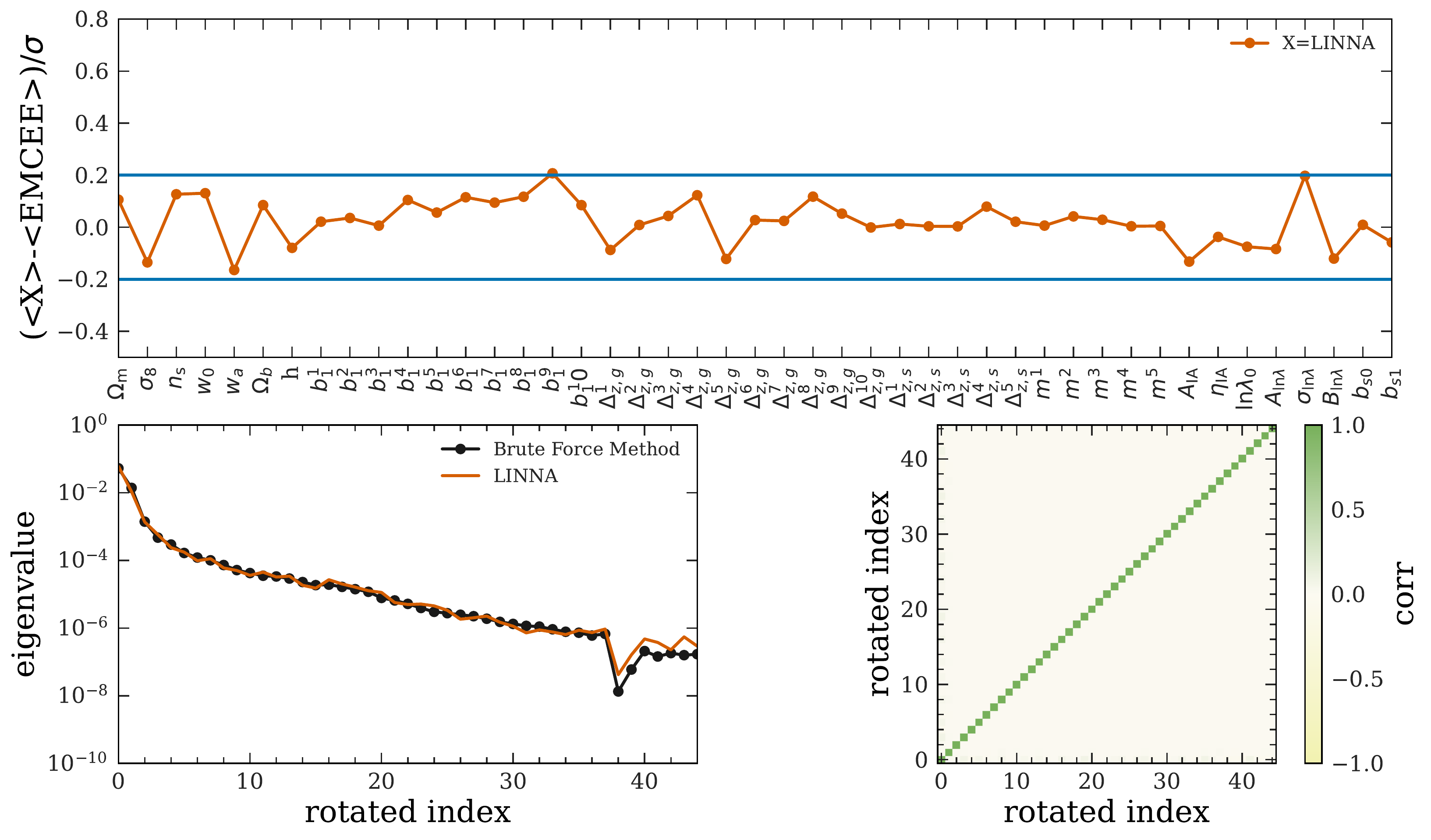}\hspace{-0.05\textwidth}
\caption{Performance of \nns{} on the \allcomb{} analysis using the simulated LSST Y10 data vector. Details of this plot are similar to figure~\ref{fig:app:multidimension_6x2pt}.}
\label{app:LSST6x2pt}
\end{figure*}

In figures \ref{app:LSST3x2pt} and \ref{app:LSST6x2pt}, we compare the result of \nns{} with the result obtained from the brute force method. We find that the level of agreement is consistent with what we found in analyzing DES Y1 data. These results indicate that \nns{} is expected to be applicable on the analysis of data with LSST Y10 precision.

\label{lastpage}
\end{document}